\documentstyle{article}
\evensidemargin 0.5cm \topmargin -1cm \input{amssym.def}
\input{amssym.tex} \newcommand{\id}{{\rm id}}\newcommand{\x}{\times}
\newcommand{\til}{\tilde} \renewcommand{\ll}{\label}
\newcommand{\be}{\begin{equation}}
\newcommand{\bea}{\begin{eqnarray}}\newcommand{\eea}{\end{eqnarray}}
\newcommand{\ee}{\end{equation}} 
\newcommand{\ci}{\cite} \newcommand{\ca}{$C^*$-algebra}
\newcommand{\rep}{representation} \newcommand{\Hs}{Hilbert space}
\newcommand{\raw}{\rightarrow} \newcommand{\law}{\leftarrow}
\newcommand{\n}{\|} \newcommand{\ot}{\otimes} 
\newcommand{\la}{\langle} \newcommand{\ra}{\rangle}
\newcommand{\inv}{^{-1}} \newcommand{\Gm}{\Gamma}
\newcommand{\dl}{\delta} \newcommand{\Dl}{\Delta}
\newcommand{\ep}{\epsilon} 
 \newcommand{\et}{\eta}
 
\newcommand{\io}{\iota} 
\newcommand{\lm}{\lambda} 
\newcommand{\rh}{\rho} \newcommand{\sg}{\sigma}
\newcommand{\Sg}{\Sigma} \newcommand{\ta}{\tau} 
\newcommand{\Ph}{\Phi} \newcommand{\phv}{\varphi}
  \newcommand{\Ps}{\Psi}
\newcommand{\A}{{\frak A}} \newcommand{\B}{{\frak B}}
\newcommand{\GC}{{\frak C}} 
 \newcommand{\CC}{{\cal C}}
\newcommand{\CE}{{\cal E}} 
\renewcommand{\H}{{\cal H}}

\newcommand{\CL}{{\cal L}} \newcommand{\C}{{\Bbb C}}
\begin{document} 
\setlength{\unitlength}{1cm}
\begin{center}
\vspace*{1.0cm} {\LARGE{\bf Compact Quantum
Groupoids}}\footnote{Dedicated to Professor H.-D.\ Doebner on the
occasion of his formal retirement. To appear in `Quantum Theory and
Symmetries' (Goslar, 18-22 July 1999), eds.\ H.-D. Doebner,
V.K. Dobrev, J.-D. Hennig and W. Luecke (World Scientific, 2000).}
\vskip 1.5cm
{\large {\bf N.P. Landsman }} 
\vskip 0.5 cm 
Korteweg-de Vries Institute for Mathematics \\ 
University of Amsterdam \\ 
Plantage Muidergracht 24 \\ 
NL-1018 TV AMSTERDAM, THE NETHERLANDS
\end{center}
\vspace{1 cm}
\begin{abstract}
Quantum groupoids are a joint generalization of groupoids and quantum
groups.  We propose a definition of a compact quantum groupoid that is
based on the theory of $C^*$-algebras and Hilbert bimodules. The
essential point is that whenever one has a tensor product over $\C$ in
the theory of quantum groups, one now uses a certain tensor product
over the base algebra of the quantum groupoid.
\end{abstract}
\vspace{1 cm} 
\section{Introduction}
Quantum groupoids are relatively novel objects in mathematics, for
which a number of distinct definitions have been proposed in the
literature.  The ``algebraic'' definition of Maltsiniotis \cite{Mal}
is relevant to Tannakian categories \ci{Del,Bru}, the ``measurable''
definition of Vallin \ci{Val} plays a role in subfactors \ci{EV}, and
the ``smooth'' definition of Lu \ci{Lu} has applications to
deformation quantization \ci{Xu} and to quantum dynamical Yang--Baxter
equations \ci{ES}. Moreover, weak $C^*$ Hopf algebras \ci{BNS}, which
occur in the algebraic theory of superselection sectors in quantum
field theory \ci{Reh}, are finite quantum groupoids in disguise
\ci{NV}.

For reasons that originate in algebraic quantum field theory, we are
here going to define compact quantum groupoids.  Our general strategy
is borrowed from noncommutative geometry \ci{Con,Wor}. Given some
geometric or topological object of ``classical'' mathematics, one
attempts to describe this object in terms of commutative \ca s endowed
with additional structure, and subsequently tries to define the
corresponding ``quantum'' object by dropping commutativity. The first
step usually involves the passage from the object in question to its
``dual'' description in terms of the space of continuous functions on
it. Ideally, this should lead to an anti-equivalence of categories,
also known as a duality.

In the simplest example, the object is a compact\footnote{We include
the Hausdorff property in the definition of compactness; in this
terminology a topological space satisfying the finite covering
property is called quasicompact.}  space $X$, which is dually and
completely characterized by the commutative unital \ca\ $C(X)$ of
continuous functions on it.  It is interesting to note that the
algebraic operation of pointwise multiplication in $C(X)$ may be seen
as a map
 $$\mu: C(X)\ot C(X)\simeq C(X\x X)\raw C(X),$$ given by $\mu
(F):x\raw F(x,x)$, which is dual to the diagonal map $\dl:x\raw x\x x$
from $X$ to $X\x X$; see \cite{Hof}.  Note that the tensor product is
not quite the algebraic one over $\C$, but a \ca ic one, which is a
certain completion of it. The latter is uniquely defined on
commutative \ca s, and satisfies $C(X)\ot C(Y)\simeq C(X\times Y)$.
The unit $I$ in $C(X)$, which is the function identically equal to
$1$, should be seen as a map $\et:\C\raw C(X)$, given by $\et(z)=zI$.
This map is dual to the map $X\raw e$, where $e=\{e\}$ is any set with
one element.

 The Gelfand-Neumark lemma on the structure of commutative \ca s actually
implies that the category of compact Hausdorff spaces is
dual to the category of commutative unital \ca s \ci{Hof}.
Hence a quantum compact space is nothing but a general \ca\ with
unit. In this case, the additional structure alluded to above merely
consist of the presence of a unit.

In this paper we extend these ideas to increasingly involved algebraic
structures, still combined with the topological structure of
compactness; this culminates in the definition of compact quantum
groupoids.
\section{Dual descriptions}
The  following diagram suggests how we may continue to add structure:

\begin{figure}[here]
\begin{center}
\begin{picture}(12,1.2)
\put(0,0){groupoids}\put(2,0){$\subset$}\put(3,0){small categories}
\put(7,0){$\subset$}\put(8.65,0){sets}
\put(0.25,1.2){groups}\put(2,1.2){$\subset$}\put(3,1.2){unital semigroups}
\put(7,1.2){$\subset$}\put(8,1.2){semigroups}
\put(0.75,0.6){$\cap$}\put(4.45,0.6){$\cap$}\put(8.85,0.6){$\cap$}
\end{picture}
\end{center}
\end{figure}

From `sets' in the lower right corner we may evidently proceed in two
directions. In one direction, one has the compact semigroups, whose
dual description is due to Hofmann \ci{Hof}. Given that a compact
semigroup $\Sg$ is dually described by $C(\Sg)$ as a topological
space, the additional structure encoding the semigroup law on $\Sg$ is
now given by a coproduct $\Delta: C(\Sg)\raw C(\Sg)\ot
C(\Sg)$\footnote{Here and in what follows, any map between two unital
\ca s is a unital $\mbox{}^*$-homomorphism as a standing assumption
(except when some conflicting property is explicitly stated, as in the
case of a coinverse). These are the arrows in the category of all
unital \ca s. }.  Explicitly, the coproduct is given by \be \Dl
f(x,y)=f(xy). \ll{coas1} \ee Associativity of the semigroup
multiplication law is expressed by the coassociativity property
imposed on the coproduct, viz.  \be ({\rm
id}\ot\Dl)\circ\Dl=(\Dl\ot{\rm id})\circ\Dl. \ll{A21} \ee

The presence of a unit $e$ in a semigroup $\Sg$ is dually encoded by a
counit $\ep: C(\Sg)\raw \C$, given by $\ep(f)=f(e)$.  The axiom
$xe=ex=x$ for all $x\in \Sg$ is dually expressed by \be ({\rm
id}\ot\ep)\circ\Dl=(\ep\ot{\rm id})\circ\Dl={\rm id}.  \ee The
singleton $e$ to which the unit in $C(\Sg)$ is dual may now be
identified with the unit in $\Sg$.

This leads to an anti-equivalence between the category of compact
semigroups (with unit) and the category of commutative unital
$C^*$-algebras with coproduct (and counit).  Although this clearly
suggests a definition, a general theory of quantum (unital) semigroups
remains to be developed.

The passage from unital semigroups $\Sg$ to groups $G$ is effected by
adding an inverse $x\raw x\inv$. In the dual description, the inverse
is encoded by the coinverse (inferiorly called antipode) $S:C(G)\raw
C(G)$, given by \be Sf(x)=f(x\inv). \ll{inverse} \ee Here, in the
commutative case, this map happens to be an involutive
$\mbox{}^*$-homomorphism, but in general a coinverse is by definition
an algebra anti-homomorphism satisfying \be (S\circ
*)^2=\id. \ll{antipode} \ee Further properties of the inverse are
dually encoded by the axioms \be \mu\circ(\id\ot S)\circ\Dl=
\mu\circ(S\ot \id)\circ\Dl=\et\circ\ep. \ll{orco} \ee Since these
axioms cannot be generalized to groupoids, for later use we here list
two well-known consequences of (\ref{orco}), viz.\ \bea S\circ\et& = &
\et ; \ll{Seta} \\ \ta\circ(S\ot S)\circ\Dl & = & \Dl\circ
S. \ll{ginv} \eea Here $\ta$ is the flip map $\ta(f\ot g)=g\ot f$.

Thus one is led to a categorical duality between compact groups and
commutative unital $C^*$-algebras with coproduct, counit, and
coinverse satisfying (\ref{orco})
\footnote{These are often called (commutative) Hopf \ca s.
Woronowicz's definition of a compact quantum group \cite{Wor} is not
based on the above axioms, but on a rather subtle reformulation of
them.}. The existence of a Haar measure, here seen as a linear map $P:
C(G)\raw \C$ satisfying \be (\id\ot P)\circ
\Dl=(P\ot\id)\circ\Dl=\et\circ P, \ll{toberep} \ee is a consequence of
these axioms. However, as a warmup for the dual description of a
compact groupoid, we note that compact groups are equally well dual to
commutative unital $C^*$-algebras with coproduct, coinverse satisfying
(\ref{Seta}) and (\ref{ginv}), and Haar measure.  Hence one can trade
a coinverse satisfying (\ref{orco}) and a counit for a coinverse
satisfying (\ref{Seta}) - (\ref{ginv}) and a Haar measure\footnote{In
their recent definition of a locally compact quantum group, Kustermans
and Vaes \cite{KV} require the existence of a Haar system as an axiom,
too.}.

In the other direction in the diagram, we may turn form sets to small
categories\footnote{A category is called small when the space of
arrows is a set.}.  A category $\CC$ consists of an arrow (or
morphism) space $\CC_1$, an object space $\CC_0$, source and target
projections $s:\CC_1\raw \CC_0$ and $t:\CC_1\raw \CC_0$, an object
inclusion map $\io:\CC_0\raw\CC_1$, and a multiplication $m:\CC_2\raw
\CC_1$. Here \be \CC_2=\CC_1\mbox{}_s *_t
\CC_1=\{(x,y)\in\CC_1\x\CC_1\mid s(x)=t(y)\}.  \ee
  
These are subject to axioms  discussed below.
In order to pass to the
dual description of a category, we assume that $\CC_1$ and $\CC_0$ are
compact. In that case, $C(\CC_0)$
and $C(\CC_1)$ are commutative \ca s with unit.
Instead of a single unit $\et:\C\raw  C(X)$, we now have two maps
$\et_t: C(\CC_0)\raw C(\CC_1)$ and $\et_s: C(\CC_0)\raw C(\CC_1)$, given
by the pullbacks $\et_t=t^*$ and $\et_s=s^*$. 

We dualize the diagram
$\CC_0\stackrel{t}{\leftarrow}\CC_1\stackrel{s}{\rightarrow}\CC_0$, by
constructing a bimodule\footnote{ An $\A-\B$ bimodule, where $\A$ and
$\B$ are complex algebras, is a vector space $\CE$ with a left action
of $\A$ and a right action of $\B$ which commute. We write
$\A\raw\CE\law\B$. } $C(\CC_0)\raw C(\CC_1)\law C(\CC_0)$. The left
action $\lm$ and the right action $\rh$ of $C(\CC_0)$ on $C(\CC_1)$
are given, for $\til{f}\in C(\CC_0)$, $g\in C(\CC_1)$, by \be
\lm(\til{f})g=\et_t(\til{f})g; \:\:\: \rh(\til{f})g=\et_s
(\til{f})g. \ll{lra} \ee

We may, of course, dualize $m$ as a map $\til{\Dl}:C(\CC_1)\raw
C(\CC_2)$, defined by $\til{\Dl} f(x,y)=f(xy)$, where $(x,y)\in
\CC_2$. This is rather awkward, and doesn't suggest a noncommutative
generalization. However, we note the $C^*$-algebraic isomorphism \be
C(\CC_1)\ot_{C(\CC_0)}C(\CC_1)\simeq C(\CC_2), \ll{iso} \ee where the
left-hand side is the completed bimodule tensor product. This is
defined by first taking the \ca ic tensor product of $C(\CC_1)$ with
itself, which is isomorphic to $C(\CC_1\x \CC_1)$, and subsequently
taking the quotient of the latter by the closed ideal generated by
$\et_s^{(1)}(C(\CC_0))-\et_t^{(2)}(C(\CC_0))$; the notation means that
$(\et_s^{(1)}\til{f})(x,y)=\til{f}(s(x))$ and
$(\et_t^{(1)}\til{f})(x,y)=\til{f}(t(y))$.  The isomorphism
(\ref{iso}) is easily established with the aid of the
Stone--Weierstrass theorem.  Accordingly, we may dualize groupoid
multiplication by a ``coproductoid''\footnote{For a related concept in
pure algebra, see Deligne \ci{Del}; also cf.\ \cite{Mal} for
applications to algebraic quantum groupoids, and \ci{Val} for a
similar coproduct occurring in the definition of a measurable quantum
groupoid.}  \be \Dl: C(\CC_1)\raw C(\CC_1)\ot_{C(\CC_0)}C(\CC_1), \ee
defined by combining $\til{\Dl}$ with the isomorphism
(\ref{iso}). Being essentially the pullback of the continuous map $m$,
this map is a morphism of unital \ca s. In addition, note hat
$C(\CC_1)\ot_{C(\CC_0)}C(\CC_1)$ is a $C(\CC_0)-C(\CC_0)$ bimodule in
the obvious way (inheriting the left action of $C(\CC_0)$ on the first
factor and the right action on the second). Now the groupoid axioms
$t(fg)=t(f)$ and $s(fg)=s(g)$ correspond to $\Dl$ being a morphism of
bimodules. As in the case of semigroups, associativity of the groupoid
product is equivalent to coassociativity of $\Dl$, now expressed by
\be ({\rm id}\ot_{C(\CC_0)}\Dl)\circ\Dl=(\Dl\ot_{C(\CC_0)}{\rm
id})\circ\Dl. \ll{A2} \ee Here the bimodule tensor product of two maps
is well defined, since each is a morphism of bimodules.

The object inclusion map $\io$ is evidently dualized by the counit
$\ep:C(\CC_1)\raw C(\CC_0)$, given by $\ep=\io^*$. Regarding
$C(\CC_0)$ as a $C(\CC_0)-C(\CC_0)$ bimodule in the obvious way, the
groupoid axiom $t\circ\io=s\circ\io={\rm id}$ is equivalent to $\ep$
being morphism of bimodules. The axiom $x\io(s(x))=\io(t(x))x=x$ for
all $x\in\CC_1$ is dually expressed by \be ({\rm
id}\ot_{C(\CC_0)}\ep)\circ\Dl=(\ep\ot_{C(\CC_0)}{\rm id})\circ\Dl={\rm
id}; \ee cf.\ the comment following (\ref{A2}).

Combining these structures, one easily obtains a duality between the
category of compact categories (with functors as arrows) and the
category of pairs of unital commutative $C^*$-algebras $(\A,\B)$,
equipped with injections $\et_{s,t}:\B\raw\A$, a counit
$\ep:\A\raw\B$, and a coproductoid $\Dl:\A\raw \A\ot_{\B}\A$.

A groupoid $\Gm$ is a small category in which every arrow is
invertible; it follows that $xx\inv=\io(t(x))$ and $x\inv
x=\io(s(x))$.  The inverse is dually described by a coinverse
$S:C(\Gm_1)\raw C(\Gm_1)$, still given by (\ref{inverse}). We now have
the property $S\circ \et_{s,t}=\et_{t,s}$, stating that $S$ is an
anti-morphism of the bimodule $C(\Gm_0)\raw C(\Gm_1)\law C(\Gm_0)$.
Since condition (\ref{orco}) cannot be written down for groupoids, we
will generalize (\ref{ginv}), and, instead of (\ref{orco}), impose \be
\ta\circ(S\ot_{C(\Gm_0)} S)\circ\Dl=\Dl\circ S. \ll{ginvoid} \ee
Because of the presence of $\ta$, the left-hand side is well defined.

There exists an analogue of Haar measure for (locally) compact
groupoids; for each $q\in \Gm_0$ one now has a measure $\mu^t_q$ on
$\Gm_1$ with support $t\inv(q)$, such that the family is
left-invariant in the obvious sense \ci{Ren,Lan}. Such a ``Haar
system'', whose existence, in contradistinction to the group case,
needs to be postulated (except for Lie groupoids, where it is
automatic \cite{Lan}), leads to a map $P: C(\Gm_1)\raw C(\Gm_0)$,
given by $P(f):q\raw \int_{t\inv(q)} d\mu^t_q(x)\, f(x)$.  This map
intertwines the left action $\lm$ of $C(\Gm_0)$ on $C(\Gm_1)$ with the
natural left action of $C(\Gm_0)$ on itself. This renders the property
\be (\id\ot_{C(\Gm_0)} P)\circ\Dl=P, \ll{Paxiom} \ee which replaces
(\ref{toberep}), well defined; strictly speaking, the right-hand side
should be preceded by the canonical isomorphism
$C(\Gm_1)\ot_{C(\Gm_0)}C(\Gm_0)\raw C(\Gm_1)$.

We eventually arrive at a duality between compact groupoids and pairs
of unital commutative $C^*$-algebras $(\A,\B)$, equipped with
injections $\et_{s,t}:\B\raw\A$, a coproductoid $\Dl:\A\raw
\A\ot_{\B}\A$, a coinverse satisfying (\ref{ginvoid}), and a Haar
measure.  The latter replaces the counit in the axiomatic
setup\footnote{In the context of finite groupoids this possibility was
first mentioned in \cite{Yam}; also see \cite{Val}.}.
\section{Intermezzo on Hilbert bimodules}
As we have seen, the dual description of categories and groupoids, as
compared with unital semigroups and groups, respectively, is that
tensorproducts over $\C$ tend to be replaced by tensorproducts over
$C(\CC_0)$ or $C(\Gm_0)$.  In defining a compact quantum groupoid, one
would now like to replace $C(\Gm_1)$ and $C(\Gm_0)$ by arbitrary
unital \ca s $\A$ and $\B$.  Our construction of a suitable bimodule
tensor product over $\B$ is based on the theory of Hilbert bimodules,
which generalize Hilbert spaces and $C^*$-algebras. We briefly review
this theory.

First, recall the concept of a Hilbert module (alternatively called a
$C^*$-module \ci{Con} or a Hilbert $C^*$-module \ci{Lance,RW,Lan})
over a given \ca\ $\B$. This is a complex linear space $\CE$ equipped
with a right action of $\B$ on $\CE$ and a compatible $\B$-valued
inner product, that is, a sesquilinear map $\langle \,
,\,\ra_{\B}:\CE\x\CE\raw\B$, linear in the second and antilinear in
the first entry, satisfying $\langle \Ps,\Ph\ra_{\B}^* =
\langle\Ph,\Ps\ra_{\B}$. One requires $\langle\Ps,\Ps\ra_{\B} \geq 0$,
and $\langle\Ps,\Ps\ra_{\B} = 0$ iff $\Ps=0$.  The space $\CE$ has to
be complete in the norm $\n\Ps\n^2= \n\langle\Ps,\Ps\ra_{\B}\n$. The
compatibility condition is $\langle\Ps,\Ph B\ra_{\B} =
\langle\Ps,\Ph\ra_{\B} B$.

For example, a Hilbert space is a Hilbert module over $\C$, and a \ca\
$\B$ may be seen as a Hilbert module over itself, in which $\langle
A,B\ra_{\B} =A^*B$. Note that the $C^*$-norm in $\B$ coincides with
its norm as a Hilbert module because of the $C^*$-axiom $\n A^*A\n=\n
A\n^2$.

A map $A:\CE\raw\CE$ for which there exists a map $A^*:\CE\raw\CE$
such that $\la \Ps,A\Ph\ra_{\B}=\la A^*\Ps,\Ph\ra_{\B}$ is called
adjointable.  An adjointable map is automatically $\C$-linear,
$\B$-linear, and bounded.  The adjoint of an adjointable map is
unique, and the map $A\mapsto A^*$ defines an involution on the space
$\CL_{\B}(\CE)$ of all adjointable maps on $\CE$. This space thereby
becomes a \ca.

An $\A-\B$ Hilbert bimodule, where $\A$ and $\B$ are \ca s, is now
defined as a Hilbert module $\CE$ over $\B$, along with a
$\mbox{}^*$-homomorphism of $\A$ into $\CL_{\B}(\CE)$. Hence one has a
space with a $\B$-valued inner product and compatible left $\A$ and
right $\B$-actions\footnote{ Note that an $\A-\B$ Hilbert bimodule is
an $\A-\B$ bimodule, since $\CL_{\B}(\CE)$ commutes with the right
$\B$-action.}, where it should be remarked that the left and right
compatibility conditions are quite different from each
other\footnote{One sometimes calls $\CE$ a $C^*$-correspondence
between $\A$ and $\B$.  Correspondences between von Neumann algebras
\ci{Con} are a special case of $C^*$-correspondences; see \ci{BDH}.}.

A Hilbert bimodule over $\B$ is a $\B-\B$ Hilbert bimodule.  For
example, a Hilbert space is a Hilbert bimodule over $\C$, and a \ca\
$\B$ is a Hilbert bimodule over itself.  One may construct a tensor
products $\CE_1\hat{\ot}_{\B}\CE_2$ of two Hilbert bimodules over
$\B$, yielding an object of the same kind.  The definition is a
special case of the following construction, which goes back to Rieffel
\ci{Rie} (also see \ci{Lance,RW}).

Given an $\A-\B$ Hilbert bimodule $\CE_1$ and a $\B-\GC$ Hilbert
bimodule $\CE_2$, define a $\GC$-valued inner product on the algebraic
tensor product $\CE_1\ot\CE_2$ by \be \la
\Ps_1\ot\Ps_2,\Ph_1\ot\Ph_2\ra^{\ot}_{\GC}=
\la\Ps_2,\la\Ps_1,\Ph_1\ra_{\B}\Ph_2\ra_{\GC}. \ll{PsBcPs} \ee This is
positive semidefinite; the completion of the quotient of
$\CE_1\ot\CE_2$ by the null space of $\la\, ,\,\ra^{\ot}_{\GC}$ is
$\CE_1\hat{\ot}_{\B}\CE_2$ as a vector space\footnote{ This space
turns out to be a certain completion of the algebraic bimodule tensor
product $\CE_1\ot_{\B}\CE_2$, so that $\hat{\ot}_{\B}$ is a
topological tensor product in the sense of Grothendieck.}. The crucial
point is that $\CE_1\hat{\ot}_{\B}\CE_2$ inherits the left action of
$\A$ on $\CE_1$, the right action of $\GC$ on $\CE_2$, and also the
$\GC$-valued inner product (\ref{PsBcPs}), so that
$\CE_1\hat{\ot}_{\B}\CE_2$ itself becomes an $\A-\GC$ Hilbert
bimodule.  For Hilbert bimodules over $\B$, simply take $\A=\GC=\B$.
\section{Corepresentations}
As a first application, we employ the topological bimodule tensor
product $\hat{\ot}_{\B}$ to formulate a theory of corepresentations of
compact groupoids. Recall \ci{Lan} that an action of a groupoid $\Gm$
on a fibered space $S\stackrel{p}{\raw}\Gm_0$ is a map $\Gm_1
\mbox{}_s *_p S\raw S$, such that $p(x\sg) = t(x)$, $\io(p(\sg))\sg =
\sg$, and $x(y\sg) = (xy)\sg$ whenever defined. Hence $x\in\Gm_1$ maps
$p\inv(s(x))$ into $p\inv(t(x))$.  For example, groupoid
multiplication is seen to be an action of $\Gm$ on
$\Gm\stackrel{t}{\raw}\Gm_0$.  Or choose $S=\Gm_0$ and $p={\rm id}$.
The product of two actions on $S_i\stackrel{p_i}{\raw}\Gm_0$, $i=1,2$,
is defined on $S_1\mbox{}_{p_1}*_{p_2}S_2 \stackrel{p}{\raw}\Gm_0$,
and is given by $x:(\sg_1,\sg_2)\mapsto (x\sg_1,x\sg_2)$.
 
A unitary \rep\ of $\Gm$ is an action for which $S$ is a continuous
field $\H$ of Hilbert spaces (cf.\ \ci{RW}) over $\Gm_0$, and
$x:\H_{s(x)}\raw \H_{t(x)}$ is a unitary operator $U(x)$\footnote{In
category language, this is simply a functor from $\Gm$ to the category
of Hilbert spaces with partial isometries as arrows.}. The tensor
product of two such \rep s is defined on the tensor product of the two
fields in the obvious way.

The dual description of unitary \rep s and their tensor products is a
generalization of the corresponding situation for unitary group \rep
s, where a \rep\ $U(G)$ on a \Hs\ $\H$ (here seen as a field of \Hs s
over a point) is dualized by a so-called co\rep\ $\dl:\H\raw \H\ot
C(G)\simeq C(G,\H)$, where $\dl\Ps(x)=U(x)\Ps$ for $\Ps\in\H$. For
compact groupoids we should regard $\Gm_1\mbox{}_t *_p\H$ as a bundle
over $\Gm_1$, with space of continuous sections
$C(\Gm_1,\Gm_1\mbox{}_t *_p\H)$.  A \rep\ $U(\Gm)$ on $\H$ then
corresponds to a co\rep\ $\dl:C(\Gm_0,\H)\raw C(\Gm_1,\Gm_1\mbox{}_t
*_p\H)$ that dualizes $U$, given by \be \dl\Ps(x)=U(x)\Ps(s(x)).  \ee

For a neat description, first note that, for $X$ compact, the class of
continuous fields $\H$ of \Hs s over $X$ precisely corresponds to the
class of full\footnote{A Hilbert module $\CE$ over $\B$ is called full
when $\la\CE,\CE\ra_{\B}$ is dense in $\B$.} Hilbert bimodules
$C(X,\H)$ over $C(X)$; indeed, this \ca ic Serre--Swan theorem is a
good way to define a continuous fields of \Hs s \ci{RW}. Secondly, we
may turn $C(\Gm_1)$ into a $C(\Gm_0)-C(\Gm_1)$ Hilbert bimodule by
putting $\lm(\til{f})g=t^*(\til{f})g$, $\rh(f)g=fg$, and (canonically)
$\la f,g\ra_{C(\Gm_1)}=\overline{f}g$ for $f,g\in C(\Gm_1)$ and
$\til{f}\in C(\Gm_0)$.  Hence we may form the tensor product of the
$C(\Gm_0)-C(\Gm_0)$ Hilbert bimodule $C(\Gm_0,\H)$ with the
$C(\Gm_0)-C(\Gm_1)$ Hilbert bimodule $C(\Gm_1)$, which yields the
isomorphism \be C(\Gm_0,\H)\hat{\ot}_{C(\Gm_0)}C(\Gm_1)\simeq
C(\Gm_1,\Gm_1\mbox{}_t *_p\H) \ll{ass} \ee as $C(\Gm_0)-C(\Gm_1)$
Hilbert bimodules. Accordingly, we may define a co\rep\ of $\Gm$ as a
map $\dl:\CE\raw \CE\hat{\ot}_{C(\Gm_0)}C(\Gm_1)$, where $\CE$ is some
full Hilbert bimodule over $C(\Gm_0)$. The defining conditions of a
groupoid actions may be dualized by the axioms \bea
\la\dl\Ps,\dl\Ph\ra_{C(\Gm_1)}= & = & s^*\la\Ps,\Ph\ra_{C(\Gm_0)};
\ll{cc1} \\ (\id\ot_{C(\Gm_0)}\Dl)\circ\dl & = &
(\dl\ot_{C(\Gm_0)}\id)\circ\dl; \\ \ll{cc2}
(\id\ot_{C(\Gm_0)}\ep)\circ\dl & = & \id.  \eea Note that (\ref{cc1})
implies that $\dl(\til{f}\Ps)=s^*(\til{f})\dl(\Ps)$ for $\til{f}\in
C(\Gm_0)$, which renders (\ref{cc2}) well defined.

Finally, the tensor product of two unitary \rep\ of $\Gm$ on
$\CE_i=C(\Gm_0,\H_i)$, $i=1,2$, is precisely defined on
$\CE_1\hat{\ot}_{C(\Gm_0)}\CE_2\simeq C(\Gm_0,\H_1\ot\H_2)$.  This
reinforces our general philosophy of replacing $\ot=\ot_{\C}$ by
$\hat{\ot}_{C(\Gm_0)}$ in passing from groups to groupoids.
\section{A definition of compact quantum groupoids}
We are now in a position to define a compact quantum groupoid. The
basic structure is a pair of unital \ca s $\A$ and $\B$ that
generalize $C(\Gm_1)$ and $C(\Gm_0)$, respectively. The injective
unital maps $\et_s:\B\raw\A$ and $\et_t:\B\raw\A$ are now required to
be a $\mbox{}^*$-homomorphism and a $\mbox{}^*$-antihomomorphism,
respectively, whose images in $\A$ commute.  The noncommutative Haar
measure, whose existence is now taken as an axiom replacing the
counit, is defined as a faithful completely positive map $P:\A\raw\B$
satisfying $P(A\et_s(B))=P(A)B$. In other words, $E=\et_s\circ
P:\A\raw\et_s(\B)\subset\A$ is a faithful conditional expectation.

This setup enables us to define a certain Hilbert bimodule $\A^-$ over
$\B$, as follows.  We first define a left $\B$-action $\lm$ and a
right $\B$-action $\rh$ on $\A$ by $\lm(B)A=A\et_t(B)$ and
$\rh(B)A=A\et_s(B)$. Subsequently, we put a $\B$-valued inner product
on $\A$ by $\la A,C\ra_{\B}=P(A^*C)$\footnote{This ``localizes'' the
canonical Hilbert module over $\A$ with respect to $P$; see
\ci{Lance}.}.  Since $P$ is faithful, we may define a new norm on $\A$
by $\n A\n^2=\n P(A^*A)\n_{\B}$.  The completion of $\A$ in this norm
is $\A^-$, which is a Hilbert module over $\B$.  If the technical
condition $\lm(\B)\subseteq \CL_{\B}(\A^-)$ is satisfied\footnote{This
is true for all von Neumann algebras and all unital commutative \ca
s. The condition may well be superfluous.}, then the given data even
define $\A^-$ as a Hilbert bimodule over $\B$.

The next step is to form the tensor product
$\A^-\hat{\ot}_{\B}\A^-$. By the general theory, this is a Hilbert
bimodule over $\B$. In addition, one obtains two
$\mbox{}^*$-homomorphisms $\phv^i:\A\raw
\CL_{\B}(\A^-\hat{\ot}_{\B}\A^-)$, $i=1,2$, by noting that the maps
$\phv_1(A): A_1\ot A_2\mapsto (AA_1)\ot A_2$ and $\phv_2(A): A_1\ot
A_2\mapsto A_1\ot(AA_2)$ from $\A$ into $\CL(\A\ot\A)$\footnote{Here
$\A\ot\A$ is the algebraic tensor product over $\C$.}  quotient and
extend to $\A^-\hat{\ot}_{\B}\A^-$.

We finally define the unital \ca\ $\A\ot_{\B}\A$ as the \ca\ in
$\CL_{\B}(\A^-\hat{\ot}_{\B}\A^-)$ that is generated by $\phv^1(\A)$
and $\phv^2(\A)$. The definition of a compact quantum groupoid is
completed by postulating a coproductoid $\Dl:\A\raw \A\ot_{\B}\A$, as
well as a coinverse $S:\A\raw\A$ that is an algebra and bimodule
anti-homomorphism.  The compatibility axioms are \bea
(\id\ot_{\B}\Dl)\circ\Dl & = & (\Dl\ot_{\B}\id)\circ\Dl; \ll{A2q} \\
(\id\ot_{\B} P)\circ\Dl & = & P; \ll{Paxiomq} \\ \ta\circ(S\ot_{\B}
S)\circ\Dl & = & \Dl\circ S; \ll{ginvoidq} \eea cf.\ (\ref{A2}),
(\ref{Paxiom}), and (\ref{ginvoid}), respectively.  One may check that
in the commutative case these axioms reduce to the dual description of
a compact groupoid\footnote{One might define a compact quantum
category by omitting the coinverse from our definition of a compact
quantum groupoid, but in the commutative case one wouldn't necessary
recover a category.}.

The reformulation of the \rep\ theory of a compact groupoid $\Gm$ as a
theory of co\rep s of $C(\Gm_1)$ on full Hilbert bimodules over
$C(\Gm_0)$ in the preceding section has been motivated by the
possibility of setting up a co\rep\ theory of compact quantum
groupoids. Indeed, one may now define a co\rep\ as a map
$\dl:\CE\raw\CE\hat{\ot}_{\B}\A$, where $\CE$ is a Hilbert bimodule
over $\B$, satisfying axioms resembling those stated for the
commutative case. The collection of all co\rep s of a given compact
quantum groupoid over $\B$ then becomes a tensor category under
$\hat{\ot}_{\B}$.
\begin{footnotesize}
  
\end{footnotesize}
\end{document}